\documentclass{ws-procs9x6-cpt19}

\usepackage{siunitx}

\begin{document}

\newcommand{\refeq}[1]{(\ref{#1})}
\def\etal {{\it et al.}}

\title{Matter Wave Interferometry for Inertial Sensing and Tests of Fundamental Physics}

\author{D.\ Schlippert,$^1$ C.\ Meiners,$^1$ R.J.\ Rengelink,$^1$ C.\ Schubert,$^1$ D.\ Tell,$^1$ \'E.\ Wodey,$^1$ \\ K.H.\ Zipfel,$^1$ W.\ Ertmer,$^1$ and E.M.\ Rasel$^1$}

\address{$^1$Institut f\"ur Quantenoptik, Leibniz Universit\"at Hannover, Welfengarten 1, \\30167 Hannover, Germany}

\begin{abstract}
Very Long Baseline Atom Interferometry (VLBAI) corresponds to ground-based atomic matter-wave interferometry on large scales in space and time, letting the atomic wave functions interfere after free evolution times of several seconds or wave packet separation at the scale of meters.
As inertial sensors, e.g., accelerometers, these devices take advantage of the quadratic scaling of the leading order phase shift with the free evolution time to enhance their sensitivity, giving rise to compelling experiments. 
With shot noise-limited instabilities better than \SI{1e-9}{m/s^2} at \SI{1}{s} at the horizon, VLBAI may compete with state-of-the-art superconducting gravimeters, while providing absolute instead of relative measurements.
When operated with several atomic states, isotopes, or species simultaneously, tests of the universality of free fall at a level of parts in $10^{13}$ and beyond are in reach. 
Finally, the large spatial extent of the interferometer allows one to probe the limits of coherence at macroscopic scales as well as the interplay of quantum mechanics and gravity.
We report on the status of the VLBAI facility, its key features, and future prospects in fundamental science.
\end{abstract}

\bodymatter

\section{Introduction}
Nearly half a century after the seminal observation of gravity-induced phase shifts on matter waves,\cite{Colella1975PRL} coherent control in atom interferometers is now a standard technique used around the world to perform inertial measurements\cite{Peters1999Nature}, determine fundamental constants,\cite{Rosi2014Nature,Parker2018Science} and to test the laws of fundamental physics.\cite{Jaffe2017NaturePhys}
At a given phase noise the intrinsic sensitivity of an atom accelerometer in the common Mach-Zehnder-type configuration\cite{Kasevich1991PRL} scales with the leading order phase shift,
\begin{equation}
\Delta\phi=\vec{k}_{\text{eff}}\cdot\vec{a}\,T^2
\end{equation}
where $\hbar \vec{k}_{\text{eff}}$ is the momentum transferred during atom-light interaction, $\vec{a}$ is the acting acceleration, and $2T$ is the wave packets' free evolution time. One can therefore increase the scale factor of the instrument $k_{\text{eff}}T^2$ by exploiting its linear scaling in atomic recoil $\hbar k_{\text{eff}}$ but also the quadratic dependency on the free evolution time $T$.
To that end, besides initiatives for operation in microgravity,\cite{Aguilera2014CQG,Becker2018Nature,Woerner2018IAC} increasing the free fall distance of ground-based devices towards long free fall tubes is a path pursued by groups around the world.\cite{Overstreet2018PRL,Zhou2011GRG}
Indeed, combined with an exquisite control over external fields and other deteriorating effects, extending the baseline of gravimeters from tens of centimeters to several meters opens the way for competition with state of the art superconducting gravimeters or quantum tests of the universality of free fall (UFF) at an unprecedented level,\cite{Hartwig2015NJP} competitive with those achieved by the best classical methods.\cite{Touboul2017PRL,Hofmann2018CQG,Schlamminger2008PRL}
In these proceedings we report on the status and key features of the Very Long Baseline Atom Interferometry (VLBAI) facility implemented in the Hannover Institute of Technology (HITec) of Leibniz Universit\"at Hannover. 
\section{The VLBAI facility}
\subsection{Design}
The VLBAI facility in Hannover is currently in its final stage of construction and consists of three main components:
\begin{enumerate}
    \item 
    At the heart of the device is a \SI{10}{m} long, vertically oriented ultra-high vacuum tube with a \SI{10}{cm} diameter in clearance for the atom optics light fields.
    In order to shield the experiment from magnetic stray fields mimicking inertial forces, the tube is enclosed in a high-performance dual layer magnetic shield reducing magnetic field gradients below $\SI{10}{nT/m}$.
    Additionally, the full baseline is equipped with a network of temperature probes for detecting and subsequently correcting errors due to temperature gradients.\cite{Haslinger2017NaturePhys}
    \item 
    Both ends of the vacuum tube will be equipped with dual-species sources\footnote{As such, the facility can be operated either in drop mode~($2\,T=\SI{0.8}{s}$) or with atoms being launched~($2\,T=\SI{2.8}{s}$).} providing the operator with (near-)quantum degenerate ensembles of stable bosonic and fermionic ytterbium isotopes as well as \textsuperscript{87}Rb.\cite{Hartwig2015NJP}
    Making use of hybrid magnetic and optical trapping techniques as well as delta-kick collimation,\cite{Muentinga2013PRL} we anticipate an atom flux on the order of $10^6\,$at/s with temperatures in the picokelvin regime.\cite{Loriani2019NJP}
    \item
    For absolute measurements, we will utilize a seismic attenuation system~(SAS) to suspend a retro reflection mirror that serves as the atom interferometer's inertial reference.
    Based on geometric anti-springs,\cite{Bertolini1999NuclInstrum} our SAS features a passive resonance frequency of hundreds of millihertz and provides means of 6-degrees-of-freedom active stabilization via electromagnetic actuation using the signals of on-board broadband seismometers as well as novel opto-mechanical devices.\cite{Richardson2019arXiv}
\end{enumerate}
The VLBAI facility aims to use rubidium, which is well-established as a standard choice for inertial sensors using readily available source and laser technology, as well as the heavy lanthanide ytterbium which offers its own benefits.
Indeed, as an effective two-electron system it has broad lines to apply strong cooling forces, as well as narrow intercombination transitions with a low Doppler-limit.
Furthermore, the bosonic isotopes all share the property of having a vanishing first order magnetic susceptibility, making systematic effects and environmental decoherence mechanisms easier to control.
Finally, the internal composition of our species offers enhanced sensitivity when searching for new physics beyond the standard model.\cite{Kostelecky2011PRD,Hohensee2013PRL}
\subsection{Performance estimation}
With the SAS and the magnetic shield tackling the dominant external noise sources, we expect an improvement in absolute gravimetry beyond the state of the art. The performance of the device will however ultimately be limited by quantum projection, i.e., shot noise. In a simple drop configuration with \SI{2e5}{at} per cycle, \SI{3}{\second} preparation time, and a free evolution time $2T = \SI{800}{\milli\second}$, the shot noise-limited sensitivity of the VLBAI facility is \SI{1.7}{\nano\meter\per\second\squared} at \SI{1}{\second}. 
In a more advanced configuration, launching \SI{1e6}{at} per cycle to reach a free evolution time of $2T = \SI{2.8}{\second}$ and using 4-photon atom optics yields a shot noise-limited acceleration sensitivity of \SI{40}{\pico\meter\per\second\squared} at \SI{1}{\second}.
For a gradiometric configuration~\cite{Rosi2014Nature,Dimopoulos2008PRD} with a baseline $L=\SI{5}{\meter}$ between the two atom interferometers with \SI{1e5}{at} per cycle, \SI{3}{\second} preparation time, and $2T = \SI{800}{\milli\second}$, the shot noise-limited sensitivity is \SI{5e-10}{\per\second\square} at \SI{1}{\second}.

A simultaneous comparison measurement with \textsuperscript{87}Rb and \textsuperscript{170}Yb has the perspective for testing the UFF\cite{Hartwig2015NJP} by determining the E\"otv\"os ratio between the two species to better than one part in $10^{13}$.
\subsection{Fundamental physics}
Beyond these metrological goals, the apparatus will also serve as a test bed for interferometry with very large scale factors as necessary for gravitational wave detection\cite{Dimopoulos2008PRD} and is intrinsically sensitive to fundamental decoherence mechanisms\cite{Bassi2017CQG} and limitations of the quantum superposition principle.\cite{Kovachy2015Nature}
\section{Conclusion \& Outlook}
The VLBAI facility will enable highly sensitive absolute gravimetry and tests of the Universality of Free Fall. It therefore offers a world-wide unique environment, both for cutting edge inertial sensing and to test our understanding of General Relativity, quantum mechanics and their interplay, possibly leading to a future reconciliation of the two.\cite{Hartwig2015NJP}
\section*{Acknowledgments}
This project is supported through the CRCs 1128~``geo-Q'', 1227~``DQ-mat'', ``Nieders\"achsisches Vorab'' in the ``Quantum- and Nano-Metrology (QUANOMET)'' initiative.
D.S. gratefully acknowledges funding by the Federal Ministry of Education and Research~(BMBF) through the funding program Photonics Research Germany under contract number 13N14875.

\end{document}